\documentclass[amsmath,amssymb,aps,reprint,superscriptaddress,longbibliography]{revtex4-1}
\usepackage{graphicx}
\usepackage{dcolumn}
\usepackage{bm}

\newcommand\p{\partial}

\newcommand\z{\zeta}

\newcommand\ta{\theta}

\newcommand\eps{\epsilon}

\newcommand\di{\textrm{d}}
\newcommand\ex{\textrm{e}}

\begin{document}

\preprint{APS/123-QED}

\title{A simple model of Keratocyte membrane dynamics  }

\author{Mohammad Abu Hamed}
\affiliation{Department of Mathematics, Technion - Israel Institute of Technology, Haifa 32000, Israel }
\affiliation{Department of Mathematics, Al-Qasemi, Academic College Of Education, Baqa El-Garbiah 30100, Israel}
\affiliation{Department of Mathematics, The College of Sakhnin - Academic College for Teacher Education, Sakhnin 30810, Israel }

\author{Alexander A. Nepomnyashchy}
\affiliation{Department of Mathematics, Technion - Israel Institute of Technology, Haifa 32000, Israel }

\begin{abstract}
We perform an analytical investigation of the cell interface dynamics in
the framework of a minimal phase field model of cell motility suggested in \cite{Ziebert2012}, which consists of two
coupled evolution equations for the order parameter and a two-dimensional vector field
describing the actin network polarization (orientation). We derive a closed evolutionary integro-differential equation governing
the cell interface dynamics. The equation includes the normal velocity of the
membrane, curvature, volume relaxation, and a parameter  that is
determined by the non-equilibrium effects in the cytoskeleton. This equation can be
simplified to obtain a Burgers-like equation. A condition on the system
parameters for the existence of a stationary cell shape is obtained.
\end{abstract}

\maketitle

\section{Introduction}

Understanding the mechanics of cell motility is needed for design of
artificial cells \cite{Kendall2018}, as well as the description of
the collective motion of biomimetic microcapsules colonies \cite{Kolmakov2010},
\cite{Kolmakov2012}, that could be applied for the design of self-healing
materials \cite{Toohey2007}. Therefore, the development of a simple
mathematical model
describing the cell interface dynamics, which is the subject of the
present
paper, can contribute to all those goals \cite{Andersson2007}.

Modeling of Eukaryotic cell motility in general and keratocyte cell in
particular, has attracted much attention in the last decades
\cite{Mogilner2008}. There are several approaches that where developed in
order to model and describe the cell crawling on the substrate
\cite{Ziebert2016}. The most common of them are the free boundary models
\cite{Nickaeen2017} and the phase field approaches \cite{Shao2010},
\cite{Camley2017}, where one considers the influence of the microscopic
subcellular elements (miosin \cite{Kimpton2015} and actin
monomers), on the cell-scale geometry changes such as protrusion and
retraction, that result in cell crawling.

In the present paper we apply the model developed by Aranson and
co-workers \cite{Ziebert2012}, \cite{Ziebert2014}, which includes the phase field
description of the cell geometry coupled with a two-dimensional vector
field of the actin network polarization.
Relative to other phase field approaches, this model can be consider as a simple
minimal model describing the cell motility \cite{Ziebert2016}.

Recently, the same model has been used in \cite{Reeves2018} in order to describe lamellipodium waves dynamics in cells, where it was
found that the cell interface dynamics could be described by the Burgers-like equation. It should be noted that in the course of the
derivations, the diffusion effects, which constitute a very basic component of the model, were ignored. In the present paper, we perform a more
rigorous analysis of the problem and preserve the diffusion effects. We obtain a more general and exact description
of the interface dynamics (\ref{gfd}) that can be reduced to a Burgers-like equation in a certain limit, thus recovering the result of
\cite{Reeves2018}.

In the remarkable work of Keren \textit{et al.} \cite{Keren2008}, the
authors find phenomenological relations between the motility at the molecular
level, within the actin network, and the overall cell geometry change,
based on experimental observation of a large number of cells. One of these relations is
 the force-velocity relation for the actin network.
The authors state that the exact form of that relation is unknown, and it
has to be found phenomenologically. In the present paper, in the framework
of the above-mentioned model, we suggest an answer to that challenging question by introducing the nonlocal equation
(\ref{gfd}), which provides the relation between the normal velocity,
membrane curvature, cell volume variation and molecular parameters of the system.

The structure of the paper is as follows: In Sec. II we present the
minimal phase field model. In Sec. III we investigate the dynamics of the circular
shape interface. In Sec. IV we consider the general shape interface. We
derive a closed evolutionary nonlocal equation that describes the interface dynamics. Also, we investigate
the stability of the circular shape front. In Sec. V we consider a special
limit where we approximate the interface dynamics by a Burgers-like
equation. Finally, in Sec. VI we present the conclusions.

\section{Formulation of the model}
We use the following simplified version of the phase-field
model suggested for the description of self-polarization
and motility of keratocyte fragments in \cite{Ziebert2012} and applied to
a wider range of problems in \cite{Ziebert2014} and \cite{ِAranson2016},
\begin{subequations}\label{Mod}
\begin{eqnarray}
&& u_t = D_u \nabla^2 u  -(1-u)(\delta-u)u -\alpha \nabla u \cdot
\textbf{P}, \label{Mod1} \\
&& \delta(x,y,t) = \delta_0 + \mu \left[ \int u(x,y,t) \di x \di y -V_0  \right] -\sigma |\textbf{P}|^2,\label{Mod2}\\
&& \textbf{P}_t = D_p \nabla^2 \textbf{P}  - \tau_1^{-1}\textbf{P}  -\beta
\nabla u, \quad |\textbf{P}|<1,\label{Mod3}
\end{eqnarray}
\end{subequations}
with boundary conditions
\begin{subequations}\label{Mod5}
\begin{eqnarray}
 && u=1, \quad  |\textbf{P}|=0,\quad \text{at} \  r\rightarrow 0 \label{Mod5.1}\\
 && u=0, \quad  |\textbf{P}|=0,\quad \text{at} \  r\rightarrow\infty \label{Mod5.2}
\end{eqnarray}
\end{subequations}
where $0<u(x,y,t)<1$ is the order parameter that tends to $1$ inside the cell
and $0$ outside, and $\textbf{P}(x,y,t)$ is the two-dimensional polarization
vector field representing the actin orientations. The model contains
several constant parameters: $D_u$ determines the width of the
diffuse interface, $D_p$ describes the diffusion of $\textbf{P}$, $\alpha$
characterizes the advection of $u$ along $\textbf{P}$,
$\beta$ describes the creation of $\textbf{P}$ at the interface,
$\tau^{-1}$ is the degradation rate of $\textbf{P}$ inside the cell,
$V_0$ is the initial overall area of the cell, $\mu$ is the
stiffness of the volume constraint, and parameter $\sigma$ is related to
the stress due to the contraction of the actin filament bundles \cite{Ziebert2012}. Notice that the model
\eqref{Mod}-\eqref{Mod5} is nonlocal due to the definition of $\delta$.

It is convenient to present the model described above in polar coordinates:
\begin{equation*}\label{}
  u=u(r,\ta,t), \quad \textbf{P}= p(r,\ta,t)\hat{r} +  q(r,\ta,t)\hat{\ta}.
 \end{equation*}
Then equations \eqref{Mod}-\eqref{Mod5} take the form,
\begin{subequations}
\begin{eqnarray}
&&u_t=D_u\nabla^2u-(1-u)(\delta-u)u- \nonumber\\
&& \alpha\left(u_rp+\frac{1}{r}u_\ta q\right),\label{pMod1}\\
&& \delta(r,\ta,t) = \delta_0 + \mu \left[ \int u(r,\ta,t) r\di r \di \ta -V_0  \right] - \nonumber\\
&& \sigma ( p^2 + q^2), \label{pMod2}\\
&& p_t = D_p \left( \nabla^2 p - \frac{1}{r^2}p-\frac{2}{r^2} q_\ta \right)- \tau^{-1} p - \beta u_r,\\
&& q_t = D_p \left( \nabla^2 q - \frac{1}{r^2}q + \frac{2}{r^2} p_\ta \right)- \tau^{-1} q - \frac{\beta}{r} u_\ta. \label{pMod5}
\end{eqnarray}
\end{subequations}
We are interested in the solution where the variable $u$ varies smoothly between $0$ at infinity (i.e., outside the
cell) and the value close
to $1$ at $r=0$ (inside the cell). We define the cell's interface $r=\rho(\ta,t)$ such
that $u(\rho(\ta,t),\ta,t)=1/2$ and assume that it is a single valued function. Recall that
$$\nabla^2  = \p_{rr} + \frac{1}{r}\p_r + \frac{1}{r^2}\p_{\ta\ta}.$$

In the next section, we start our analysis with consideration of the circular
cell dynamics.

\section{Circular cell dynamics}

Let us consider a cell of a circular shape. All the fields have an axial
symmetry, i.e., $u=u(r,t)$,
$\textbf{P}=p(r,t)\hat{r}$, and $\rho=\rho(t)$.

In the present paper, the basic assumption is that the thickness of the cell wall (i.e., the width of the
transition zone, where $u(r,t)$ is changed from nearly 1 to nearly 0), is
small compared to the size of the cell.
In that case, the nonlocal term in \eqref{Mod2} or \eqref{pMod2} can be estimated as
 \begin{equation*}\label{}
   \int u(x,y,t) \di x \di y \approx 2\pi\int_{0}^{\rho(t)} u(r,t) r\di
r\approx \pi \rho^2 (t).
 \end{equation*}
The model determined by equations \eqref{pMod1}-\eqref{pMod5} takes the following local form,
\begin{subequations}
\begin{eqnarray}
&& u_t = D_u (u_{rr} + \frac{1}{r} u_r ) -(1-u)(\delta-u)u -\alpha  u_r p,\label{rMod1} \\
&& \delta(r,t) = \delta_0 + \mu \pi( \rho^2 (t) -\rho_0^2  ) -\sigma p^2,\label{rMod2}\\
&& p_t = D_p \left( p_{rr} + \frac{1}{r}p_r -\frac{1}{r^2}p \right) -
\tau_1^{-1} p  -\beta  u_r ,\label{rMod3}\\
&& u_r(r=0)=0, \quad u(r\rightarrow\infty)=0, \quad 0<u<1,\label{rMod4}\\
&& p(r=0)=p(r\rightarrow\infty)=0, \quad |p|<1.\label{rMod5}
\end{eqnarray}
\end{subequations}
Also, by definition,
\begin{equation*}\label{eq6}
u(\rho(t),t)=\frac{1}{2}.
\end{equation*}
As mentioned above, while the width of the cell wall is $O(1)$, the cell radius $\rho(t)$ is large. Let us rescale the
cell radius,
\begin{equation}\label{eq7}
\rho(t)=\epsilon^{-1}R(t), \quad R(t)=O(1),
\end{equation}
and introduce the variable that describes the distance to the cell interface,
$$z=r-\rho(t)=O(1).$$
Then
\begin{equation}\label{eq10}
  \frac{1}{r} = \frac{\epsilon}{R(t)} - \frac{\epsilon^2 z}{R^2 (t)} +...
\end{equation}

Also, we assume that function $\delta(r,t)$ is always close to $1/2$, i.e., $\delta(r,t)=1/2+O(\epsilon)$ in order to utilize-the Ginzburg Landau theory, and parameter
$\alpha=O(\epsilon)$. These conditions guarantee that the variables in the transition zone are changed slowly, which
allows to rescale the time variable,
\begin{equation}\label{}
  \tilde{t}=\epsilon^2t=O(1).
\end{equation}
We define,
\begin{equation}\label{eq8}
  R(t) = \tilde{R}(\tilde{t}) , \quad u(r,t) = \tilde{u}(z,\tilde{t}), \quad p(r,t)=\tilde{p}(z,\tilde{t}).
\end{equation}
Parameter $\alpha$ is rescaled as
\begin{equation}\label{eq11}
\alpha=\epsilon A.
\end{equation}
For $\delta (r,t)$ we choose:
\begin{subequations}\label{delta}
\begin{eqnarray}
&& \delta (r,t) = \tilde{\delta}(z, \tilde{t} ) = \frac{1}{2} +  \tilde{\delta}_1 (z, \tilde{t} ), \label{delta1}\\
&&\tilde{\delta}_1(z,\tilde{t})=\delta_1+\pi\mu\eps^{-2} \left(\tilde{R}^2(\tilde{t})-\tilde{R}_0^2\right)-\nonumber\\
&&\sigma\tilde{p}^2=O(\epsilon).\label{delta2}
\end{eqnarray}
\end{subequations}
We see that quantities $\delta_1$, $\mu\eps^{-2}$ and $\sigma$ should be $O(\epsilon)$. That give us the scaling,
\begin{equation*}\label{eq12}
 \pi \eps^{-2}\mu = \eps M, \ \sigma = \eps S, \ \delta_1 = \eps B.
\end{equation*}
Note that unlike \cite{Ziebert2014} and \cite{Reeves2018}, we do not assume that $\delta_1=0$.

Let us substitute \eqref{eq7}-\eqref{delta} into \eqref{rMod1}-\eqref{rMod5}, using the chain rule:
\begin{equation*}
\partial_t=-\epsilon\tilde{R}_{\tilde{t}}\partial_z+\epsilon^2\partial_{\tilde{t}},\;\partial_r=\partial_z.
\label{eq11}
\end{equation*}
Later on, we drop the tildes. We obtain the system of equations that describes the dynamics in the transition zone $z=O(1)$ and
determines {\em the inner solution}:
\begin{eqnarray*}
&& -\eps R_t u_z = D_u \left(u_{zz} + \frac{\eps}{R(t)} u_z \right) - (1-u)\left(\frac{1}{2}-u\right)u -\nonumber\\
&& \epsilon (1-u)u[B+M(R^2(t)-R^2_0)-Sp^2]-\nonumber\\
&&\epsilon A u_z p + O(\eps^2),\\
&& -\eps R_t p_z = D_P \left( p_{zz} + \frac{\eps}{R(t)} p_z \right) - \nonumber\\
&& \tau^{-1} p -\beta u_z + O(\eps^2).
\end{eqnarray*}
The {\em outer solutions} of \eqref{rMod1}-\eqref{rMod5} are trivial: inside the cell, $u=1$ and $p=0$; outside the
cell, $u=p=0$. Therefore, matching the inner solution to the outer solutions, we obtain the following boundary
conditions:

\begin{eqnarray*}
&& u (z\rightarrow-\infty)=1, \quad u (z\rightarrow \infty)=0, \\
&& p (z\rightarrow \pm\infty)=0.
\end{eqnarray*}
Let us introduce the expansions
\begin{equation}\label{expan}
  u = u_0 + \eps  u_1+..., \quad p = p_0 + \eps p_1 +...
\end{equation}
At the leading order, we obtain:
\begin{subequations}
\begin{eqnarray}
&& D_u u_{0zz} = (1-u_0)\left(\frac{1}{2}-u_0 \right)u_0\label{led1}\\
&& D_p p_{0zz} -\tau^{-1}p_0 = \beta u_{0z}\label{led2}\\
&& u_0(z\rightarrow-\infty)=1, \quad u_0(z\rightarrow \infty)=0,\label{led3} \\
&& p_0(z\rightarrow \pm\infty)=0.\label{led4}
\end{eqnarray}
\end{subequations}

\begin{figure}
  \centering
  \includegraphics[scale=0.3]{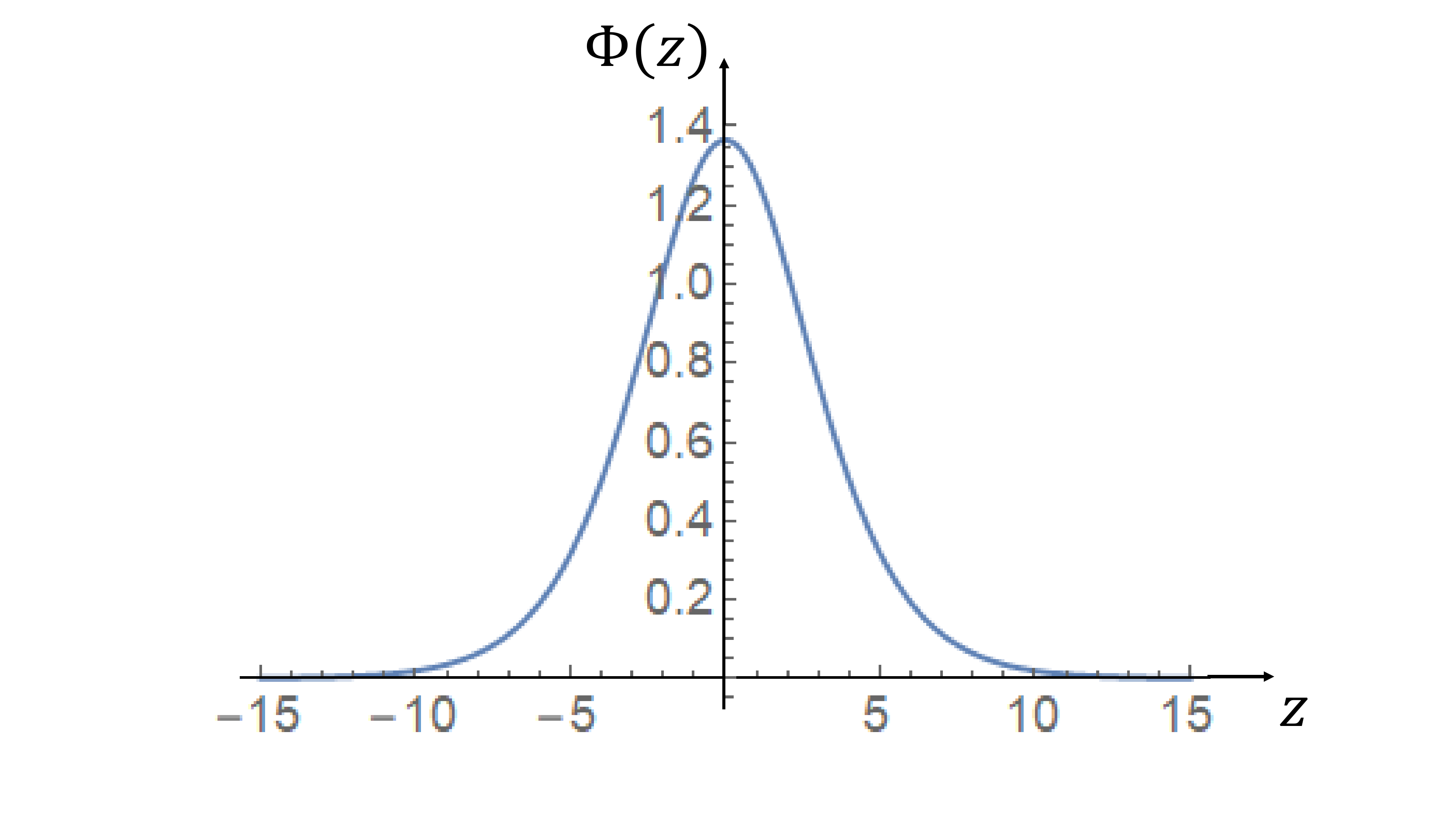}
  \caption{ The plot of the function $\Phi(z)$ that describes the polarization field in (\ref{p0(z)}), (\ref{Phi(z)}) .
We employ the values of  parameters $\tau=10, \ D_u =1, \ D_p =0.2, \ A=1, \ B=1.2, \ M=0.4, \ S=1.5, \ R_0=2. $   } \label{Phi-z}
\end{figure}

The solution of \eqref{led1} and \eqref{led3} is known from the Ginzburg-Landau theory,
\begin{equation}\label{u0(z)}
   u_0 = \frac{1}{2} \left[ 1-\tanh\left(\frac{z}{\sqrt{8D_u}}\right) \right],
\end{equation}
while equations \eqref{led2} and \eqref{led4}  can be solved via Fourier transform,
\begin{subequations}\label{Phi}
\begin{eqnarray}
  && p_0 (z) =\beta \Phi(\tau,D_u , D_p, z  ), \label{p0(z)} \\
  && \Phi(\tau,D_u , D_p, z  ) =\label{Phi(z)}\\
  &&\frac{1}{8}\sqrt{\frac{\tau}{2 D_u D_p}} \int_{-\infty}^{\infty} \ex^{-|r|/\sqrt{\tau D_p}} \cosh^{-2} \left( \frac{r- z}{\sqrt{8D_u}} \right) \di r.\nonumber
\end{eqnarray}
\end{subequations}
See Fig. \ref{Phi-z} for the plot of the function $\Phi(z)$ that is basic for our analysis.

The equation for $u_1$ at $O(\eps)$ is
\begin{eqnarray*}
&& D_u u_{1zz} - \left(\frac{1}{2} -3 u_0 + 3u_0^2 \right)u_1 =\nonumber \\
&& (1-u_0)u_0 \eps^{-1} \tilde{\delta}_1 + A u_{0z}p_0 - \frac{D_u}{R(t)}u_{0z} - R_t u_{0z}.
\end{eqnarray*}
The boundary conditions are:
$$u_1(z\to\pm\infty)=p_1(z\to\pm\infty)=0.$$

\begin{figure}
  \centering
  \includegraphics[scale=0.3]{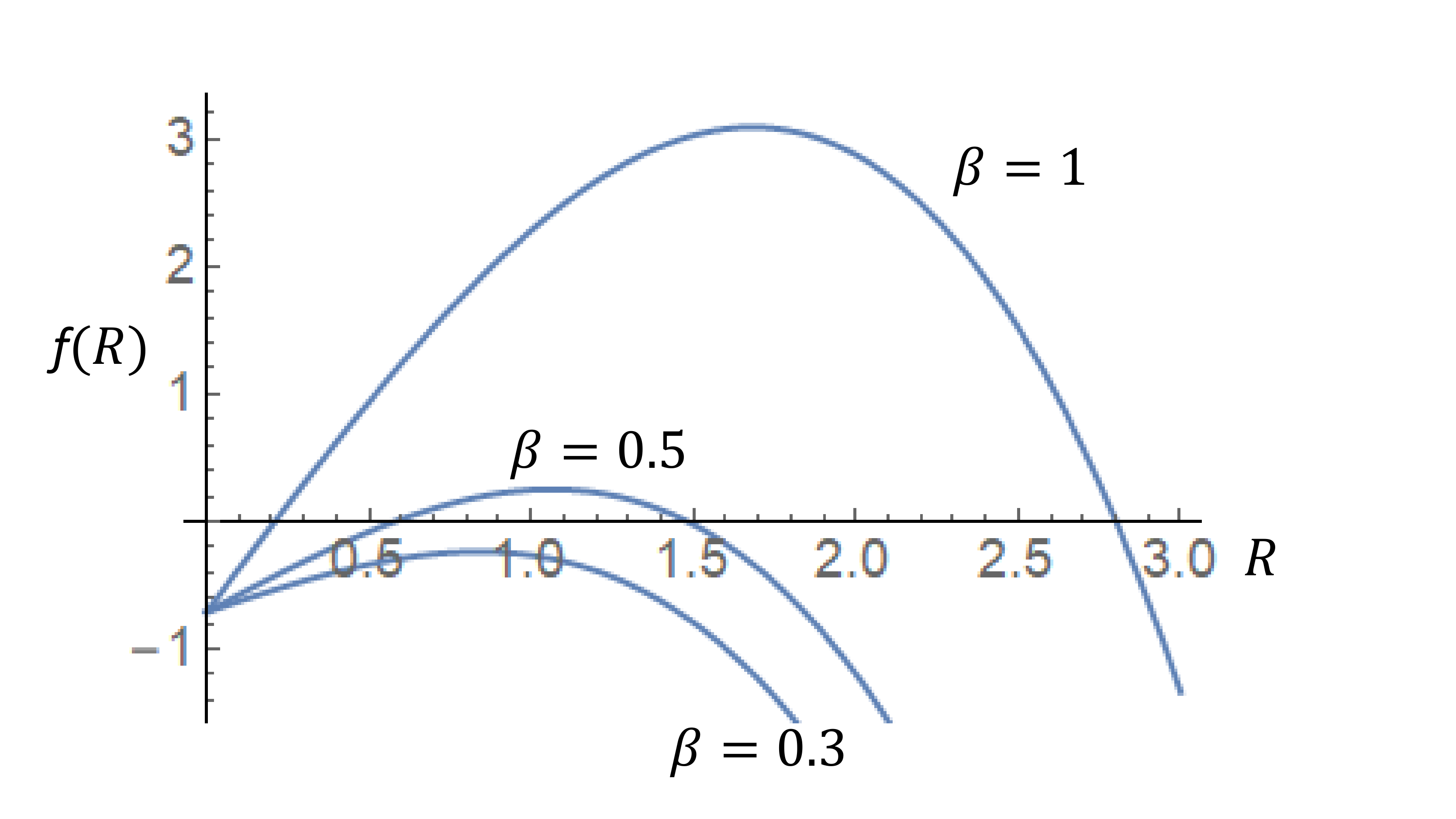}
  \caption{ Plots of the function $f(R)$ in (\ref{R(t)1}) for the values
$\beta=1,0.5$ that manifest stable and unstable states, and for the value $\beta=0.3$ that is nonphysical since it does not include any steady states.
 We use the same values of parameters as in Fig. \ref{Phi-z}.   } \label{f-R}
\end{figure}

The solvability condition yields the following closed form of the front dynamics equation,
\begin{eqnarray}
&& \int_{-\infty}^{\infty} \di z u_{0z} \Big\{ - R_t u_{0z} - \frac{D_u}{R(t)}u_{0z}+ A u_{0z}p_0 + \nonumber \\
&& u_0 (1-u_0) \left[ B + M \left( R^2 (t) -  R_0^2 \right) - S p_0^2  \right]  \Big\} =0. \label{R(t)0}
\end{eqnarray}
Expression \eqref{u0(z)} yields,
\begin{equation*}\label{}
\int_{-\infty}^{\infty} \di z u_{0z}^2=\frac{1}{6\sqrt{2D_u}}, \quad  \int_{-\infty}^{\infty} \di z u_{0z}u_0(1-u_0)=-\frac{1}{6}.
\end{equation*}
Thus, equation \eqref{R(t)0} can be written in the form
\begin{eqnarray}
&&\frac{1}{\sqrt{2D_u}}\left(R_t+\frac{D_u}{R}\right)=M[R_0^2-R^2(t)]- \nonumber\\
&&B+ \Omega(\beta),\label{R(t)}\\
&&\Omega(\beta) =  6\beta \left(A \Omega_1 -  \beta S \Omega_2  \right)\nonumber 
\end{eqnarray}
where
\begin{eqnarray*}
&& \Omega_1 (\tau ,D_u , D_p  ) = \int_{-\infty}^{\infty} \Phi(z) u_{0z}^2 \di z>0, \\
&&\Omega_2(\tau,D_u,D_p)=\nonumber\\
&&\int_{-\infty}^{\infty}\Phi^2(z)(1-u_0)u_0
u_{0z}\di z<0.
\end{eqnarray*}
It is more convenient to use the following form of equation \eqref{R(t)} governing the front dynamics,
\begin{eqnarray}
&& \frac{1}{\sqrt{2D_u}} R_t  = \frac{f(R)}{R}, \label{R(t)1}\\
&& f(R) = -MR^3 + ( MR_0^2 -B+\Omega(\beta) )R -\sqrt{\frac{D_u}{2}}, \nonumber
\end{eqnarray}
In Fig. \ref{f-R} we plot the function $f(R)$ for several values of $\beta$ for some values of parameters, these graphs show the existence of
two stationary radii, stable and unstable, for $\beta=1$ and $\beta=0.5$,
while no stationary states for $\beta=0.3$. Therefore, there exists a critical value $\beta_c$ such that there are no stationary
solutions when $\beta<\beta_c$. Below we find that $\beta_c=0.41$.

Indeed, the critical value $\beta_c$ has to satisfy three constraints: (i) $MR_0^2 +\Omega(\beta_c) -B>0$, which
guarantees the existence of maximum of $f(R)$ at a certain $R=R_{*}$, (ii) $f(R_*,\beta_c)=0$, (iii) $f'_R(R_*,\beta_c)=0$. As a
result we find that $\beta_c$ is the positive solution of the quadratic equation
  \begin{equation}\label{b-c}
  MR_0^2 +\Omega(\beta_c) -B= \frac{3}{2}\sqrt[3]{MD_u},
\end{equation}
which can be found explicitly. For values of parameters indicated in Fig. \ref{Phi-z} we fined $\beta_c=0.41$ .

\begin{figure}
  \centering
  \includegraphics[scale=0.3]{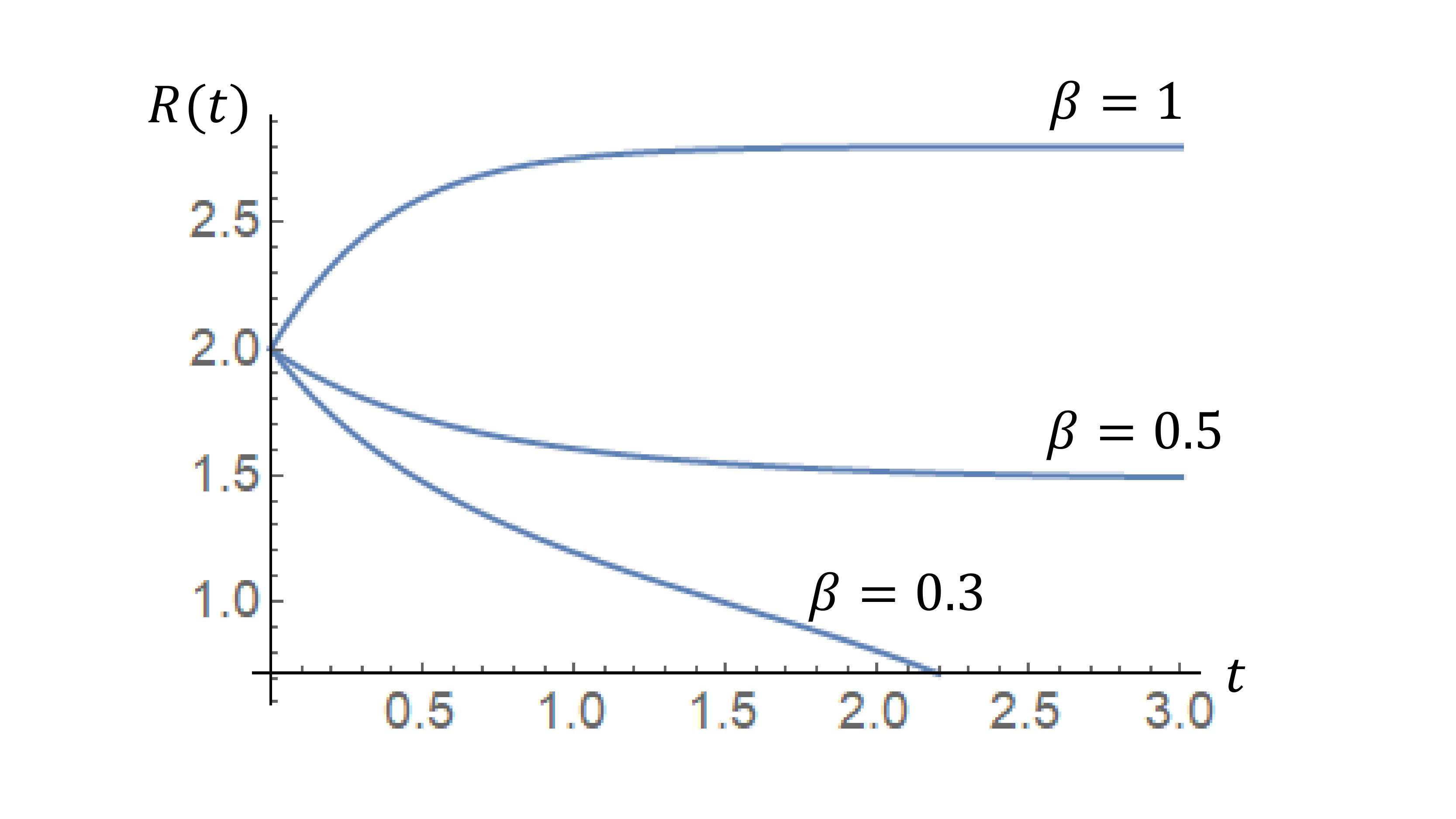}
  \caption{ Numerical solutions of the ODE (\ref{R(t)}) that yields $R(t)$
for values $\beta=1,0.5$ that manifest the existence of stationary radius,
and for the value $\beta=0.3$ that is below the critical value
$\beta_c=0.41$ that is meaningless in the context of cell dynamics.  We use the same values of parameters as in Fig. \ref{Phi-z}.
    } \label{R-t}
\end{figure}

In Fig. \ref{R-t} we present the numerical solution of the ODE \eqref{R(t)1}
for the  values $\beta=1$ and $0.5$. As we can see, the circular front radius can increase or decrease  monotonically until it
reaches the steady state value. This is because of the volume is not conserved but influences the dynamics through the parameter
$\delta(r,t)$. For the value $\beta=0.3$ the cell shrinks until it disappears, which means that for
such values of parameters, the model does not reflect the true behavior of cells.

\section{Dynamics of general cell shape}

\subsection{Evolution equation}

In this section we consider the evolution of a non-axisymmetric cell
shape. We employ the same scaling and definitions as in the previous
section. When considering the contributions of derivatives with respect to
the azimuthal variable, we use the following expressions:
\begin{eqnarray}
&& \p_\ta \rightarrow -\eps^{-1} R_\ta \p_z + \p_\ta \nonumber\\
&& \p_\ta^2  \rightarrow \eps^{-2} R_\ta^2 \p_{z}^2 - \eps^{-1}( R_{\ta\ta}\p_z +2R_\ta \p_{z\ta}^2 ) + \p_{\ta}^2 \nonumber\\
&& \frac{1}{r}\p_\ta = -\frac{R_\ta}{R} \p_z +O(\eps), \ \ \frac{1}{r^2}\p_\ta = -\eps \frac{R_\ta}{R^2} \p_z +O(\eps^2) \nonumber\\
&& \nabla^2  = \left( 1+\frac{R_\ta^2}{R^2} \right) u_{zz} + \nonumber\\
&& \frac{ \eps }{R}\left( u_z -\frac{R_{\ta\ta}}{R}u_z -\frac{2R_\ta}{R} u_{z\ta} - \frac{2zR_\ta^2}{R^2}u_{zz}  \right)+O(\eps^2). \label{LO}
\end{eqnarray}
 We approximate the nonlocality in \eqref{pMod2} as follows,
 \begin{equation*}\label{}
   \int u(x,y,t) \di x \di y \sim   \frac{\eps^{-2}}{2} \int_{0}^{2\pi} R^2 (\ta,t)  \di \ta .
 \end{equation*}
The same expansions \eqref{expan} are applied. Define the auxiliary
function
\begin{equation*}\label{}
   \Lambda(\ta,t)= \left( 1+ \frac{R_\ta^2}{R^2} \right)^{-1/2}
 \end{equation*}
and perform the proper change of variable,
$$z = \Lambda^{-1}\zeta.$$
At the leading order we find,
 \begin{eqnarray*}\label{}
&& D_u \Lambda^{-2} u_{0zz} = (1-u_0)(\frac{1}{2}-u_0)u_0,\\
&& D_p \Lambda^{-2} p_{0zz} - \tau^{-1} p_0 = \beta u_{0z},\\
&& D_p \Lambda^{-2} q_{0zz} - \tau^{-1} q_0 = -\beta \frac{R_\ta}{R} u_{0z},
\end{eqnarray*}
therefore similarly to the previous section one can calculate the solutions
\begin{subequations}
 \begin{eqnarray}\label{}
&& u_0 (z)=   \frac{1}{2} \left[ 1-\tanh\left(\frac{\z (z) }{\sqrt{8D_u}}\right) \right], \label{l1}\\
&& p_0(z)   = \beta\Lambda  \Phi(\zeta)\label{l2}\\
&& q_0(z)   = -\beta\Lambda  \frac{R_\theta}{R}  \Phi(\zeta)\label{l3}
\end{eqnarray}
\end{subequations}
The equations for $u$ at the order $O(\eps)$ have the form,
\begin{eqnarray}\label{}
&& D_u \Lambda^{-2} u_{1zz} - \left(\frac{1}{2} -3 u_0 + 3u_0^2 \right)u_1 = - R_t u_{0z}+\nonumber\\
&& D_u \left( -\frac{u_{0z}}{R} + \frac{R_{\ta\ta}}{R^2} u_{0z} + \frac{2R_{\ta}}{R^2} u_{0z\ta} + \frac{2 z R_{\ta}^2}{R^3} u_{0zz}  \right)\label{curef}\\
&& + A  u_{0z} \left( p_0 -\frac{R_\ta}{R} q_0 \right) +\nonumber\\
&& (1-u_0)u_0 \left\{ B + M \left[ \frac{1}{2\pi} \int_{0}^{2\pi} R^2 (\ta,t)\di\ta - R_0^2 \right]-  S(p_0^2 + q_0^2)  \right\}. \nonumber
\end{eqnarray}
Denote $a=1/\sqrt{2D_u}$. While applying the solvability condition, one has to calculate the following integrals,
\begin{eqnarray}\label{}
&&\int_{-\infty}^{\infty} \di z u_{0z}\left( -\frac{u_{0z}}{R} + \frac{R_{\ta\ta}}{R^2} u_{0z} + \frac{2R_{\ta}}{R^2} u_{0z\ta} + \frac{2 z R_{\ta}^2}{R^3} u_{0zz}  \right)=\nonumber\\
&& -a\kappa(\ta,t), \quad \kappa(\ta,t)=\frac{R^2 + 2 R_\ta^2 - R R_{\ta\ta} }{( R^2 + R_\ta^2 )^{3/2}}, \label{int1}\\
&&  \int_{-\infty}^{\infty} \di z u_{0z}^2 \left( \bar{p} -\frac{R_\ta}{R} \bar{q} \right)= \beta \Omega_1, \label{int2}\\
&&  \int_{-\infty}^{\infty} \di z u_{0z} u_0 (1-u_0) (p_0^2 + q_0^2)= \beta^2 \Omega_2.\label{int3}
\end{eqnarray}
For the first integral \eqref{int1} we refer the reader to \cite{AbuHamed2016} where we calculate the same expression in details. For
the integrals \eqref{int2}, \eqref{int3}, notice that the terms dependent on $(\theta,t)$ are vanish. The solvability condition yields a
closed evolutionary equation for the front dynamics of the cell, which is an integro-differential equation, i.e., it is nonlocal, unlike that obtained
in the circular case \eqref{R(t)1},
\begin{eqnarray}
&& a \Lambda R_t = -aD_u \kappa(\ta,t) \label{gfd} \\
&& - \left\{ B + M \left[ \frac{1}{2\pi} \int_{0}^{2\pi} R^2 (\ta,t)\di\ta - R_0^2 \right]  \right\}+ \Omega(\beta),\nonumber
\end{eqnarray}
where $\kappa$ is the mean curvature. Indeed the radial dynamics \eqref{R(t)} is recovered, if $R$ does not depend on $\theta$.
By a proper scaling transformation, $t\rightarrow a^2 t $ and $R(t)\rightarrow aR(t) $, the equation of motion of the cell boundary can
be brought to a canonical form,
 \begin{equation*}\label{}
   v_n = -D_u \kappa -  \delta V + \Omega.
 \end{equation*}
 Here we denote
$$\delta V (t) = B+ M \left[ \frac{1}{2\pi} \int_{0}^{2\pi} R^2 (\ta,t)\di\ta - R_0^2 \right] . $$
Notice that in \eqref{gfd} the expression $\Lambda R_t$ corresponds to the normal velocity of the interface, thus that equation is a generalization
of the well-known curvature flow. In addition it suggests an answer for the unrevealed force – velocity
relation for the actin network that was highlighted in \cite{Keren2008}.

In Fig. \ref{fluc}, the results of the numerical simulation of the fluctuation $ \delta R = R(\ta,t)-R_0 \ll 1$, which has been carried out
using equation \eqref{gfd}, are shown. We employ the \verb"DSolve" function of Mathematica Wolfram. Notice that the fluctuation is relaxed to constant value, that means that any slightly deformed shape will relax to some circular shape, in agreement with the result of the next subsection.

\subsection{Stability of the circular cell shape}

For investigating the stability of the circular stationary front $\bar{R}(t)$ that is governed by \eqref{R(t)1} we substitute the disturbance,
\begin{equation*}\label{}
  R(\ta,t) = \bar{R}(t) + \hat{R}(\ta,t),
\end{equation*}
into equation \eqref{gfd}. The linearized equation for a small front deformation is
\begin{equation*}\label{hatR}
  \hat{R}_t = D_u \frac{\hat{R} + \hat{R}_{\ta\ta}}{ \bar{R}^2 } -
\frac{M}{3\pi a} \bar{R} \int_{0}^{2\pi} \hat{R} \di \ta,
\end{equation*}
subject to a boundary condition
$$R(\theta+2\pi,t)=R(\theta,t).$$
Let us consider the evolution of a mode
$$\hat{R}= \varphi_k(t) \ex^{ik\ta},\;k\in  \mathbb{Z},$$
with a definite azimuthal number $k$. Only for the mode with $k = 0$, which corresponds to a change of the
circle radius, the integral term in \eqref{gfd} contributes into the evolution equation for the disturbance. We find that
$$\p_t \varphi_0=\lambda(t)\varphi_0,$$
where
$$\lambda(t)=\sqrt{2D_u}\frac{d}{dR}
\left(\frac{f(R)}{R}\right)\left|_{R=\bar{R}(t)}=
\frac{D_u}{\bar{R}^2(t)}-\frac{2M\bar{R}(t)}{3\pi a}.\right.$$
For sufficiently small $\bar{R}(t)$, the disturbance $\varphi_0$ can
grow, but when $\bar{R}(t)$ approaches its stationary values, the
derivative $d(f(R)/R)/dr$ at $R=\bar{R}(t)$ becomes negative (see Fig. \ref{f-R}),
hence the circular cell is stable with respect to circular disturbances, as it was shown in the previous section.

For any non-axisymmetric normal modes, $k\neq 0$, we obtain,
 \begin{equation*}\label{}
 \p_t \varphi_k =\frac{ D_u (1-k^2) }{ \bar{R}^2 } \varphi_k.
\end{equation*}
The disturbances with $|k| = 1$ correspond to a spatial shift  of the
circle and do not change with time. Disturbances with $|k| \geq 2$, which
describe the shape distortions, decay with time. Thus, a small
deformation of the circular shape does not produce any instabilities,
in agreement with numerical study of \cite{Ziebert2012}. Note that a
finite-amplitude disturbance can generally generate an instability.

\subsection{The limit of $R_\ta^2/R^2\ll1$ and the Burgers-like equation }

As the next step, let us consider a nonlinear evolution of a small disturbance satisfying the condition $R_\ta^2/R^2\ll1$, hence $\Lambda\sim 1$.


In that case, the Laplace operator in \eqref{LO} can be approximated as
 \begin{equation}\label{LO1}
   \nabla^2 u  =   u_{zz} +   \frac{ \eps }{R}
\left( u_z -\frac{R_{\ta\ta}}{R}u_z -\frac{2R_\ta}{R} u_{z\ta}  \right)+
\textit{o}(\epsilon).
 \end{equation}
 At the leading order we obtain the same solutions $u_0(z)$ and $p_0(z)$
as in \eqref{u0(z)} and \eqref{Phi}, respectively, and also
\begin{equation*}\label{}
    q_0(z) = -\beta \frac{R_\ta}{R} \Phi(z),
 \end{equation*}
While applying the solvability condition to equation \eqref{curef}, we find that
\begin{eqnarray}
&&  \int_{-\infty}^{\infty} u_{0z}^2 \left( p_0 -\frac{R_\ta}{R} q_0 \right)\di z= \beta \left(1 + \frac{R_\ta^2}{R^2} \right)\Omega_1, \label{int4}\\
&&  \int_{-\infty}^{\infty} u_0 (1-u_0) u_{0z}  ( p_0^2 + q_0^2) \di z = \beta^2 \left(1 + \frac{R_\ta^2}{R^2} \right)\Omega_2. \label{int5}
\end{eqnarray}
Notice that when calculating \eqref{int4} and \eqref{int5} we preserve the negligible expression $R_\ta^2/R^2$ while it is omitted in the
approximation of the Laplacian \eqref{LO1}.



Finally, the dynamics of the interface is governed by the following nonlinear equation
\begin{eqnarray}\label{}
   && R_t  = D_u \left( \frac{R_{\ta\ta}}{R^2} - \frac{1}{R} \right) - \nonumber\\
   && \sqrt{2D_u}\delta V (t) +  \sqrt{2D_u} \Omega(\beta) \left(1 + \frac{R_\ta^2}{R^2} \right),\label{Bur0}
 \end{eqnarray}
with the boundary condition
$$R(\theta+2\pi,t)=R(\theta,t),$$
in agreement with \cite{Reeves2018}.

Let us consider the fluctuation $\delta R = R(\ta,t)-R_0 \ll 1$; then equation \eqref{Bur0} takes the form,
\begin{eqnarray}\label{}
   && \delta R_t  = D_u \left( \frac{\delta R_{\ta\ta}}{R_0^2} - \frac{1}{R_0} + \frac{\delta R}{R_0^2} \right) - \nonumber\\
   && \sqrt{2D_u} \delta V (t) +  \sqrt{2D_u} \Omega(\beta) \left(1 +
\frac{\delta R_\ta^2}{R_0^2} \right),\label{Bur1}
 \end{eqnarray}
with the boundary condition
$$\delta R(\theta+2\pi,t)=\delta R(\theta,t).$$
Notice that in \eqref{Bur1} we preserve leading, first-order correction
and quadratic
terms.

Applying the operator $\p_\ta$ to the both sides of equation \eqref{Bur1} and denoting $G=\p_\ta \delta R$, we obtain the following nonlinear
evolution equation,
 \begin{equation}\label{Bur2}
   \p_t G = \frac{D_u}{R_0^2} \left( \p_\ta^2 G + G +  \sqrt{\frac{2}{D_u}}
\Omega(\beta) \p_\ta G^2  \right).
 \end{equation}
The solution should satisfy the periodicity condition
$$G(\theta+2\pi,t)=G(\theta,t),$$
and the condition
$$\int_0^{2\pi}G(\theta,t)d\theta=0,$$
which follows from the periodicity of $\delta R$ and definition of $G$. The obtained equation resembles the Burgers equation, but it contains an
additional linear term that can lead to a local growth of $G$. An equation similar to \eqref{Bur2} was studied in \cite{Mikishev1993} in another
physical context.

The results of the numerical simulation of the temporal evolution of the fluctuation $\delta R$ in the frameworks of equations \eqref{Bur1} and \eqref{gfd} almost coincide, see Fig. \ref{fluc}. Therefore, equation \eqref{Bur1} may be considered as a reasonable simplification of the more rigorous and general equation \eqref{gfd}. Note however in the case when the initial curvature is large in some points the simulation revel a significant deference in the behavior of solutions of \eqref{gfd}, \eqref{Bur1} see Fig. \ref{fluc1} and \ref{fluc2}.

\section{Conclusion}
We have performed an analysis of the cell surface dynamics in the
framework of the minimal phase field model of
cell motility that was developed and investigated numerically in
\cite{Ziebert2012}, \cite{Ziebert2014}, and \cite{Ziebert2016}, where the
order parameter $u$ is coupled with polarization (orientation) vector
field $\textbf{P}$ of the actin network. We considered the
axisymmetric case (circular shape interface), where we obtained a closed
ordinary differential equation describing the evolution of the radius \eqref{R(t)1}. We
found the minimum value $\beta_c$ for the actin creation that is compatible with the existence of a stationary cell solution \eqref{b-c}. We found that
when $\beta_c < \beta$, the circular cell can have some stationary radius, while in the case $\beta\leq\beta_c$ the cell shrinks
until it disappears, which is meaningless in the context of cell dynamics. Also, we consider the general shape  cell dynamics. We found the
leading order solutions, \eqref{l1}-\eqref{l3}, and derived a closed integro-differential equation \eqref{gfd} governing the cell
dynamics, which includes the normal velocity of the membrane, curvature, volume relaxation rate, and a parameter $\Omega$ determined by
the molecular effects of the subcell level.

We found an equation of motion of the cell interface that can be written in the canonical form,
 \begin{equation*}\label{}
   v_n = -D_u \kappa -  \delta V + \Omega.
 \end{equation*}

\begin{figure}
  \centering
  \includegraphics[scale=0.3]{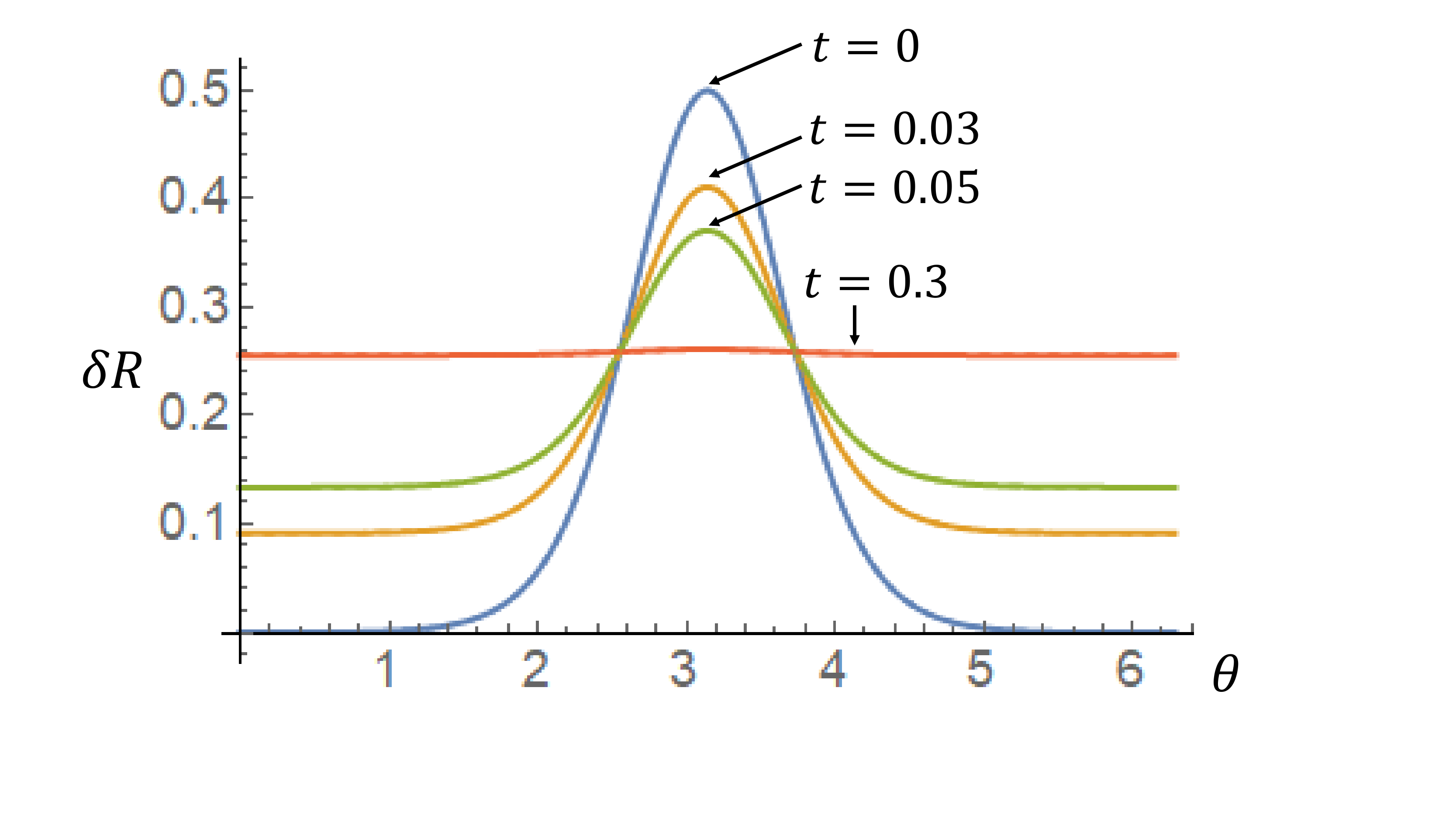}
  \caption{ Numerical solutions for the fluctuation field $\delta R =
R(\ta,t)-R_0$ of  equations   (\ref{gfd}) and (\ref{Bur1}) subject to the
initial peak perturbation $R(\ta,0)= \frac{1}{2}\cosh^{-4}(\ta-\pi)$ . Notice
that both of them coincide.     The following set of parameters were employed $\beta=1,\ \tau=10, \ D_u =1, \ D_p =0.2, \ A=1, \ B=0.2, \ M=1, \ S=1.5, \ R_0=5 $   } \label{fluc}
\end{figure}

\begin{figure}
  \centering
  \includegraphics[scale=0.3]{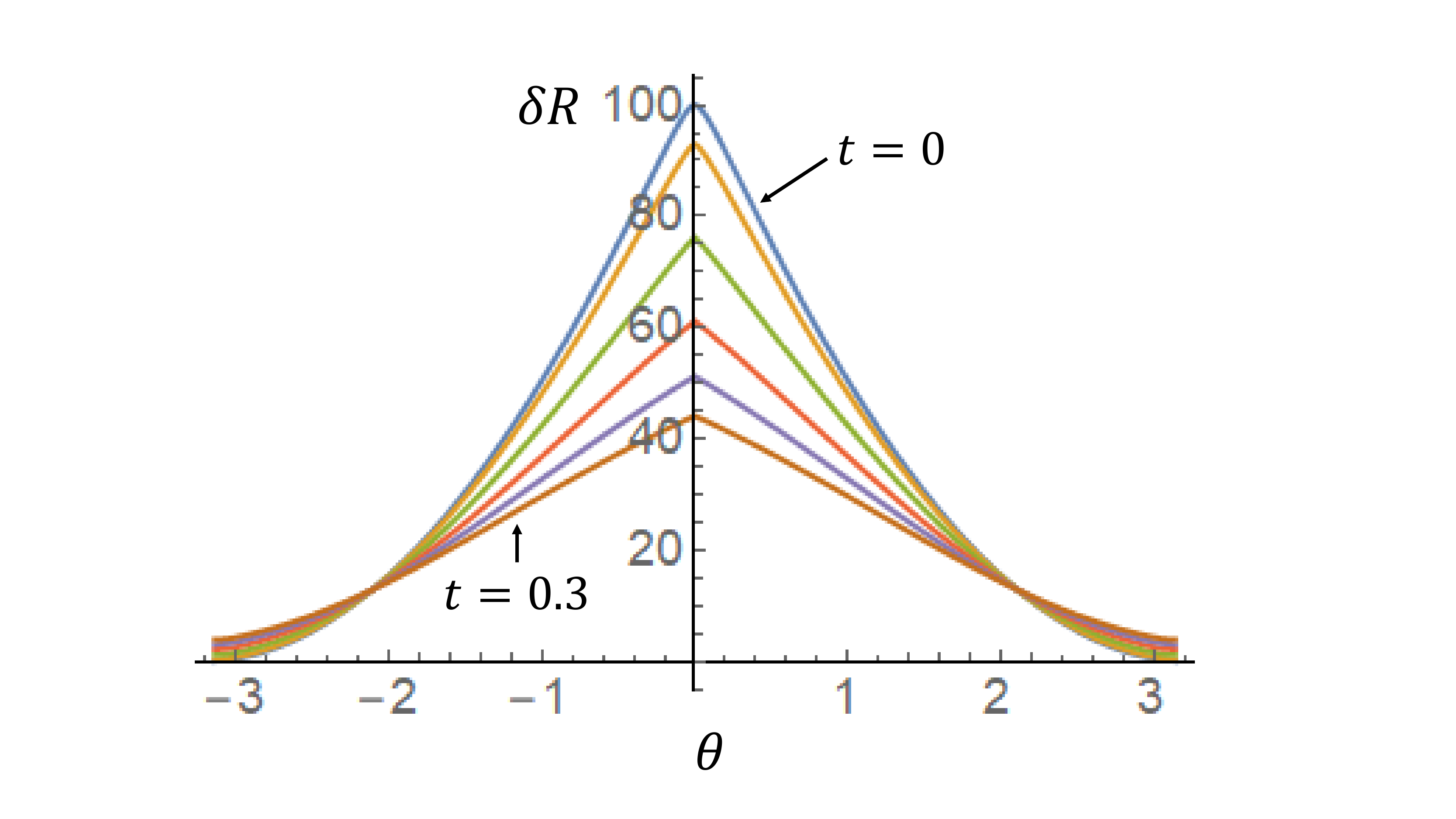}
  \caption{ Numerical solutions for the fluctuation field
  $\delta R = R(\ta,t)-R_0$ of equation (\ref{gfd}) at the time sequence
$t=0,0.05,0.1,0.15,0.2,0.4,0.6,0.8$ subject to the initial peak perturbation
  $R(\ta,0)= 100\left( \frac{\ta^2}{\pi(|\ta| + 0.1) } -1 \right)^2$ .    The following set of parameters were employed $\beta=1,\ \tau=10, \ D_u =1, \ D_p =0.2, \ A=1, \ B=0, \ M=0.01, \ S=1.5, \ R_0=5 $.  } \label{fluc1}
\end{figure}

\begin{figure}
  \centering
  \includegraphics[scale=0.3]{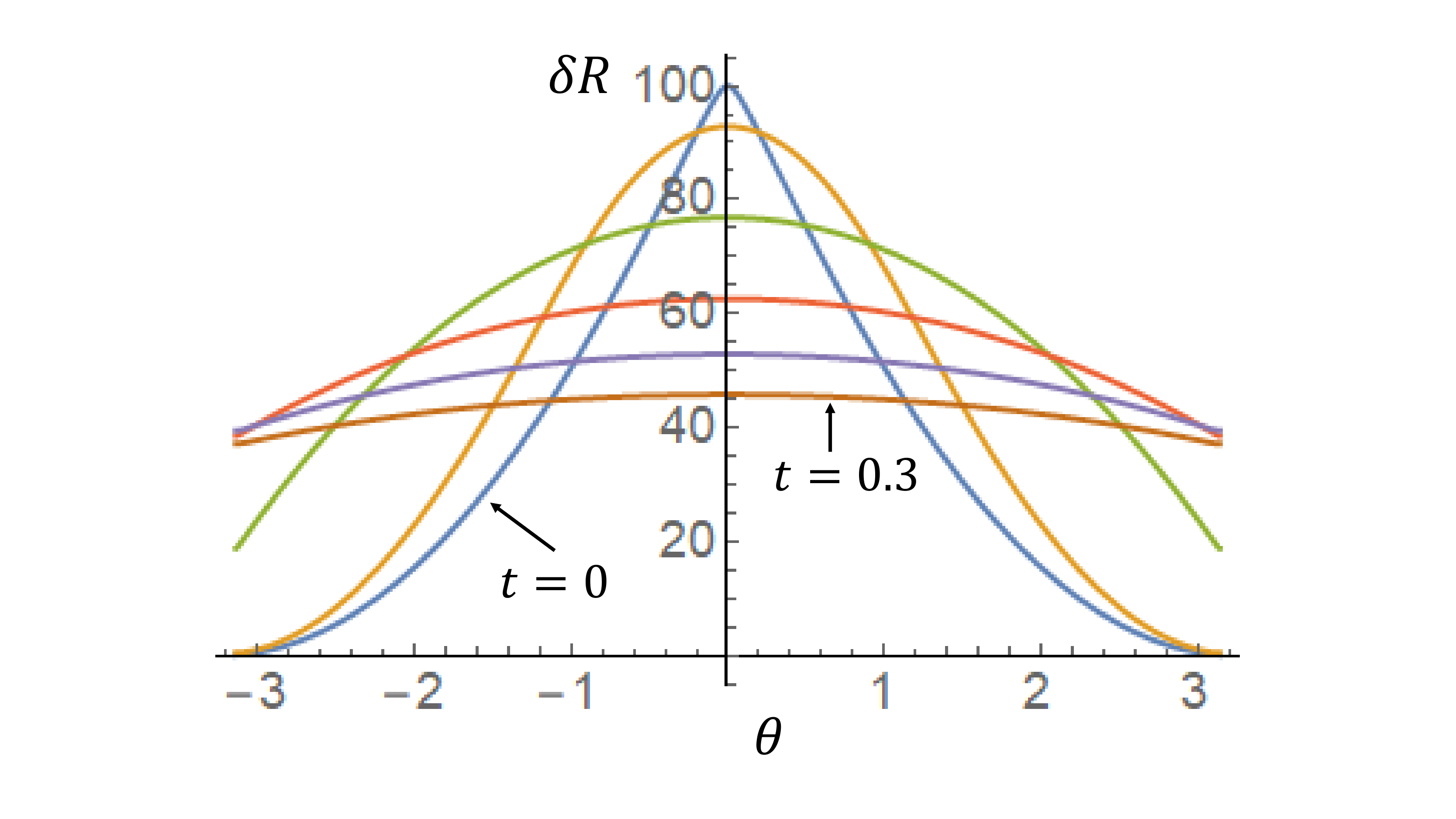}
  \caption{ Numerical solutions for the fluctuation field $\delta R =
R(\ta,t)-R_0$ of equation (\ref{Bur1}). We use the same initial peak, time sequence, and values of parameters as in Fig. \ref{fluc1}.  } \label{fluc2}
\end{figure}

Finally we considered a simplification of our main equation, \eqref{gfd}, in the limit $R_\ta^2/R^2\ll1$, and obtained
a Burgers-like equation \eqref{Bur1} for evolution of a small fluctuation $\delta R$. Fig. \ref{fluc} confirm such approximation, since the numerical solutions of the both equations \eqref{gfd} and \eqref{Bur1} almost coincide. Thus, the result of \cite{Reeves2018}, obtained with diffusion terms
neglected, is recovered from our more rigorous and precise equation \eqref{gfd}. However for other values of parameters and when the initial curvature is large in some points the simulations of equations \eqref{gfd}, \eqref{Bur1} revel a significant deference see Fig. \ref{fluc1}-\ref{fluc2}.

\bibliography{LMCM1.6.bbl}{}

\end{document}